# Photon-triplets for quantum optics generated by a phase-matched third-order difference-frequency mixing in a KTiOPO₄ bulk crystal pumped at 532 nm

GASPAR MOUGIN-TRICHON,[1] VERONIQUE BOUTOU,[1,*] CORINNE FELIX,[1] DAVID JEGOUSO,[1] BENOIT BOULANGER[1,**]

[1]*Univ. Grenoble Alpes, CNRS, Institut Néel, 38000 Grenoble, France*
*veronique.boutou@neel.cnrs.fr*

**We report implementation and modelling of an efficient photon-triplets generation experiment based on a difference-frequency-mixing of two picosecond beams at 532 nm and 1491 nm in a type II phase-matched KTP crystal. The photon-triplets flux was measured as a function of the energy of the two incident beams using a coincidence protocol. A maximal flux of 11.6 photon-triplets *per* second was achieved. These experimental data were satisfactorily described by a semiclassical model based on the quantum fluctuations of vacuum and the classical equations of nonlinear optics.**

The present study deals with the generation of photon-triplets by third-order parametric down conversion (PDC), *i.e.* $\hbar\omega_p \rightarrow \hbar\omega_{sti} + \hbar\omega_s + \hbar\omega_i$, stimulated over only one mode of the triplet, here at $\omega_{sti}$, as shown in Fig.1. This interaction is equivalent to a third-order difference-frequency-mixing between the waves at $\omega_p$ and $\omega_{sti}$ leading to the generation of two new waves at $\omega_s$ and $\omega_i$, or to an optical parametric amplification at $\omega_{sti}$.

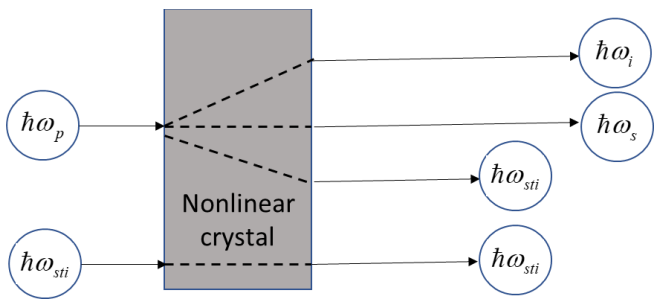

Fig. 1. Scheme of third-order parametric down conversion pumped at $\omega_p$ and stimulated at $\omega_{sti}$ leading to the generation of the photon-triplet $\{\omega_{sti}, \omega_s, \omega_i\}$.

Generating such a state of light is of prime interest in quantum optics since it allows new protocols of quantum information to be implemented [1]. The modelling of such a process cannot be carried out using the classical formalism alone since the initial conditions of the generated light at $\omega_s$ and $\omega_i$ correspond to a nil energy. In a preliminary study [2], we performed such an experiment for which we measured the triplet flux as a function of the stimulation intensity, and we used a quantum model to describe these experimental data [2,3], but the correction to apply was strong. Furthermore, in a recent work, we proposed a semi-classical model describing a collinear phase-matched second-order spontaneous PDC (SPDC) that had been experimentally validated between 1 W/cm² to 10 GW/cm² of pump intensity at 527 nm and 532 nm in a type II phase-matched bulk KTP crystal under the undepleted pump approximation (UPA) [4].

In the present work, we have optimized the experiment of photon-triplets generation and completed the set of experimental data by measuring the generated flux using a protocol of coincidences as a function of the stimulation and pump energies. We have also adapted to photon-triplets the semi-classical model previously used in the case of photon-pairs.

The classical part of the new model proposed here is based on the coupled differential equations relative to the spatial derivative of the complex amplitudes of the interacting waves, *i.e.* $\partial E_a/\partial Z$ with a = (p, sti, s, i), in the undepleted pump approximation (UPA), *i.e.* $\partial E_p/\partial Z = \partial E_{sti}/\partial Z = 0$, and in a collinear phase-matching configuration, *i.e.* $\omega_p n_p - \omega_{sti} n_{sti} - \omega_s n_s - \omega_i n_i = 0$ where $n_a$ is the refractive index at the circular frequency $\omega_a$ [5,6].

The quantum part corresponds to the quantum fluctuations of vacuum that are taken as the initial values of the amplitudes of the generated signal and idler fields, *i.e.* [7]:

$$E_{s,i}(Z=0) \equiv \Delta E_{s,i}^{\text{vacuum}} = \sqrt{\Delta\omega_{s,i} \hbar \omega_{s,i} / 4\pi c \varepsilon_0 n_{s,i} S} \quad (1)$$

where $\Delta\omega_{s,i}$ is the spectral linewidth of the generated photons, and $S$ the cross section of the overlap between the pump and stimulation beams. The three photons of the triplet at $\{\omega_{sti}, \omega_s, \omega_i\}$ being emitted simultaneously, the photon-triplet flux $N_{\text{triplets}}$ is given by the photon-pair flux at $\{\omega_s, \omega_i\}$. Then, it comes for the flux of the generated triplets:

$$N_{\text{triplets}}(L) = \frac{\varepsilon_0 n_s c S}{4\hbar\omega_s}\left[\Delta E_s^{\text{vacuum}}(\cosh(\beta L)-1) + \sqrt{\frac{\omega_s n_i}{\omega_i n_s}}\Delta E_i^{\text{vacuum}} \sinh(\beta L)\right]^2 \quad (2)$$

L is the crystal length, $\beta = \chi_{eff}^{(3)} E_{sti}(0) E_p(0) \sqrt{\kappa_s \kappa_i}$ with $\chi_{eff}^{(3)}$ the third-order effective coefficient and $\kappa_{s,i} = \omega_{s,i}/2n_{s,i}c$ : $\chi_{eff}^{(3)} = 7.8 \times 10^{-22} m^2/V^2$, $\kappa_s = 1.096 \times 10^6 m^{-1}$ and $\kappa_s = 1.047 \times 10^6 m^{-1}$ for the present experiments performed in a type II phase-matched KTP crystal pumped in the x-axis: the pump at $\lambda(\omega_p)$=532 nm is polarized along the y-axis of the crystal, the stimulation at $\lambda(\omega_{sti})$=1491 nm along the z-axis, and the signal and idler are at $\lambda(\omega_s)=\lambda(\omega_i)$=1654 nm polarized along the y-axis and z-axis, respectively [8]. By considering that $\Delta\omega_s \approx \Delta\omega_i (\equiv \Delta\omega)$ since $\omega_s = \omega_i$ and $n_i - n_s \approx 0$ corresponding to the birefringence in the considered phase-matching direction, Eq. (2) reduces to:

$$N_{triplets}(L) = \frac{\Delta\omega}{16\pi} [\exp(\beta L) - 1]^2 \qquad (3)$$

This semiclassical model will be compared to a quantum model in the Heisenberg representation based on the nonlinear momentum operator detailed in a previous work [2].

The second-harmonic of a 1064-nm-laser (Leopard, Continuum) with a pulse duration of a $\tau$ =15 ps and a repetition rate of 10Hz is used as a pump. The output of a tunable optical parametric generator OPG (TOPAS, Light conversion) pumped by the third harmonic of the same picosecond laser is fixed at $\lambda(\omega_{sti})$=1491 nm as a stimulation field. The spectral widths of the pump and stimulation fields are $\Delta\lambda(\omega_p) = 0.5\ nm$ and $\Delta\lambda(\omega_{sti}) = 3.2\ nm$, respectively i.e. $\Delta\omega_p = 3.33 \cdot 10^{12}\ rad.Hz$ and $\Delta\omega_{sti} = 2.71 \cdot 10^{12}\ rad.Hz$. Consequently, the idler and signal spectral widths equals $\Delta\omega \approx 2.489 \cdot 10^{12}\ rad.Hz$ i.e. $\Delta\lambda \approx 3.56\ nm$, which leads to quantum fluctuations of vacuum magnitude of $\Delta E_{s,i}^{vacuum} \approx 26.7\ V.m^{-1}$. Both pump and stimulation beams are carefully superimposed in time and space and focused with a 40-cm-focal lens in the 1-cm-long x-cut KTP crystal. The focal lens is chosen so that both input beams Rayleigh lengths exceed 2 cm, which is long enough regarding the crystal length in order to assume the parallel beam propagation inside the crystal. Both the pump and stimulation waist *radii* are measured using a beam profiler camera (Ophir) at the focal plane. We found $w_p = 47.5 \pm 2.5\ \mu m$ and $w_{sti} = 72.5 \pm 2.5\ \mu m$. Note that the size of the stimulation beam exceeds the size of the pump beam although the two beam axes are perfectly superimposed at the crystal. Hence, the stimulation energy is reduced by a factor $\gamma = 1 - \exp[2(w_p/w_{sti})^2] = 0.537$. Behind the KTP crystal, a set of filters (FS), including two Notch at 532nm (Thorlabs), two long pass filters at 1550 nm and one at 1535 (Semrock), two RG100 (Schott) and a band-pass filter around 1650 nm (Thorlabs), is used to block the pump and stimulation fields and any residual parasitic sources. This is mandatory in a collinear phase-matched configuration where there is no spatial splitting of the generated signal and idler beams. A zoom fiber collimator (ZFB, Thorlabs) is further used to collect and inject the two unseeded fields generated at $\lambda(\omega_s)=\lambda(\omega_i)$=1654 nm in a dedicated single fiber (SMF). In order to separate the polarization components y and z, the SMF fiber is connected to the input fiber port of a variable polarization beam-splitter bench (Thorlabs) which turns the z-polarization field propagation direction by 90° according to the y-polarization as shown in Figure 2. The y- and z-polarization fields are injected in two SMF fibers at the two output ports of the bench. Each fiber is then connected to a superconducting nanowire single photon detector (SNSPD) from IDQ with more than 80% efficiency at 1660 nm (83% for SNSPD1 and 85% for SNSPD2 at 1550 nm). The coincidence between the two SNSPDs can then be measured for different pump and stimulation energies and different integration times, knowing that detection of a pair ($\omega_s$, $\omega_i$) reveals a photon-triplets generation event. Adjusting this detection setup is very tricky. However, the global efficiency of this complete detection setup, including the filters set, the zoom fiber collimator ZFB, the SNSPD efficiency and the fiber polarization beam splitter has been measured and ranges from 2% to 20%.

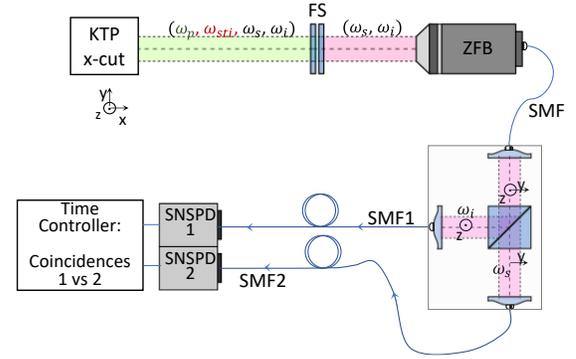

Fig. 2. Detection scheme. Top view of the setup. FS stands for Filters Set, ZFB for Zoom Fiber Collimator, SMF for Single Mode Fibers and SNSPD for Superconductor Nanowire Single Photon Detectors. At the output of the crystal, the y- and z-polarized generated fields are injected in a SMF through the ZFB and directed to the input port of a polarization beam splitter bench to be detected on SNSPD1 and SNSPD2 for coincidence measurements.

The key advantage of coincidence measurements in the context of weak photon flux is the high signal-to-noise *ratio* thanks to the use of a short coincidence time-window, *i.e.* the time duration where a detected photon on each SNSPD creates a coincidence event. Using a 100-ps-window, the noise is around $10^{-7}$ coincidence/s and can be therefore neglected. However, three major challenges are addressed in the present study. Firstly, the ability to detect true coincidence events knowing that the pulse duration is smaller than the SNSPD recovering time, meaning a maximum of one photon/pulse is detected [9]. Secondly, the probability of detection is not 100% because of losses due to the global efficiency of the detection setup. Thirdly, the ability to perform statistics with a low repetition rate, *i.e.* 10 Hz imposed by the pump laser, which leads to a weak number of events to detect. During an experiment, the measured quantity is the ratio between the number of coincidence events and the associated pulses number. Then, by considering that the two photons ($\omega_s$, $\omega_i$) of each generated triplet are deterministically separated and propagate separately towards the two SNSPDs, the average number of photon-triplets at the exit of the nonlinear medium can be expressed as [10]:

$$N_{triplets}(L) = -\frac{\ln[1 - \sqrt{\hat{\eta}(L)}]}{T_F} \qquad (4)$$

The quantity $T_F$ is the transfer function of the setup, that corresponds to the global efficiency of the detection setup, and $\hat{\eta}(L) = N_{triplets}^{measured}(L)/N_{pulse}$, where $N_{triplets}^{measured}$ is the measured coincidence events raw number while $N_{pulse}$ is the number of pulses. It is important to notice that two uncorrelated photons – for instance resulting from photons belonging to two different triplets – can simultaneously tick on the two SNSPDs, thus creating a "wrong" coincidence event. Therefore, to ensure true coincidence measurements, it will be important to verify that $N_{triplets} < 1$ triplet/pulse.

The measured observables are the energy of the pump and stimulation modes, respectively written $\xi_p, \xi_{sti}$. They are linked to the time and space gaussian electric field complex amplitudes $E_p$ and $E_{sti}$ by:

$$\xi_{p,sti} = \frac{\tau}{4}\left(\frac{\pi}{2}\right)^{3/2} \varepsilon_0 c n_{p,sti} \left(w_{p,sti} E_{p,sti}\right)^2 \quad (5)$$

where $w_0$ is the waist *radius* and $\tau$ the pulse duration.

Then, accounting for the spatial overlap between the pump and stimulation beams discussed previously and Eq. (5), the parameter $\beta$ in Eq. (3) writes:

$$\begin{cases} \beta = \gamma \chi_{eff}^{(3)} \sqrt{\dfrac{\psi_p \psi_{sti} \xi_p \xi_{sti} \kappa_p \kappa_{sti}}{w_p^2 w_{sti}^2}} \\ \psi_{p,sti} = \left(\dfrac{\pi}{2}\right)^{-3/2} \dfrac{4}{c\varepsilon_0 \tau n_{p,sti}} \end{cases} \quad (6)$$

Two sets of data have been collected, *i.e.* Set A keeping the pump energy to $\xi_p = 19.3$ µJ while varying $\xi_{sti}$ from 3.04 µJ to 11.2 µJ and Set B fixing $\xi_{sti} = 19$ µJ and $\xi_p$ varying from 8.1 µJ to 9.5 µJ. Three measurements were performed for each energy combination $\{\xi_p, \xi_{sti}\}$. For different energy combinations, the integration time can differ, ranging from 120 s to 20mn depending on the intensity stability and the value of $\hat{\eta}$. Table 1 gathers the average coincidence counts measured by the SNSPDs and the standard deviations, without any statistical correction. As discussed before, this correction, as well as the criterion $N_{triplets} < 1$ triplets/pulse, are function of the global system transfer function $T_F$ that ranges from 2% to 20%.

In a first step, the corrected experimental data are compared to the semi-classical and quantum models described previously by normalizing to the lowest energy point of each set, A and B. The corresponding curves are shown in Fig. 3 and Fig. 4. Horizontal error bars come from energy fluctuation of the laser while vertical bars come from disparities in the measurements. On both sets A and B, a strong agreement between the experimental data and the semi-classical model described by Eq. (3) is observed, especially striking for set A. In contrast, the quantum model gradually discontinues from the data and would need to be interpolated as expected from prior work [2].

**Table 1. Signal and idler coincidences without any statistical corrections as a function of the pump and stimulation energies.**

| Fixed Energies [µJ] | Tunable Energies [µJ] | Coincidences events $\hat{\eta}(L)$ [$\times 10^{-3}$/pulse] |
|---|---|---|
| **Set (A)** $\xi_p = 19.3 \pm 4.7$ | $\xi_{sti} = 3.04 \pm 0.6$ | $0.567 \pm 0.125$ |
| | $\xi_{sti} = 5.01 \pm 0.8$ | $1.77 \pm 0.125$ |
| | $\xi_{sti} = 7.4 \pm 1.3$ | $5.67 \pm 0.118$ |
| | $\xi_{sti} = 9 \pm 1.9$ | $8.78 \pm 1.40$ |
| | $\xi_{sti} = 11.2 \pm 1.6$ | $14.3 \pm 3.42$ |
| **Set (B)** $\xi_{sti} = 19 \pm 3.8$ | $\xi_p = 8.1 \pm 2.5$ | $1.33 \pm 0.656$ |
| | $\xi_p = 8.8 \pm 2.2$ | $3.03 \pm 1.24$ |
| | $\xi_p = 9.5 \pm 3.1$ | $4.33 \pm 1.65$ |

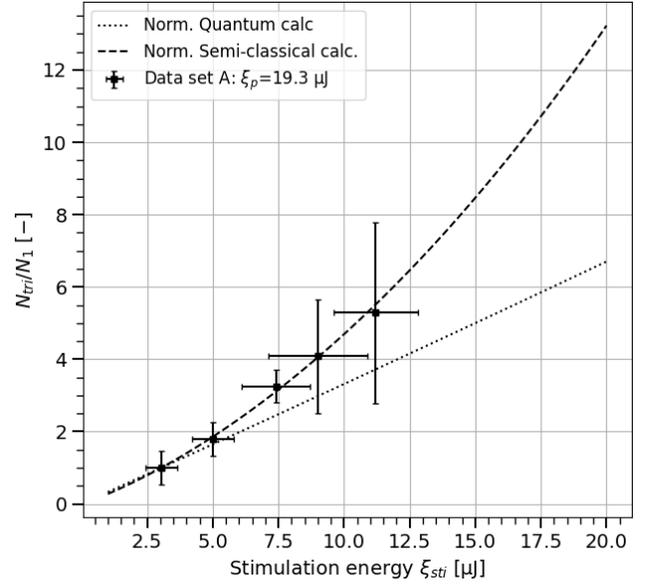

Fig. 3. Normalized flux of the generated photons as a function of the stimulation energy.

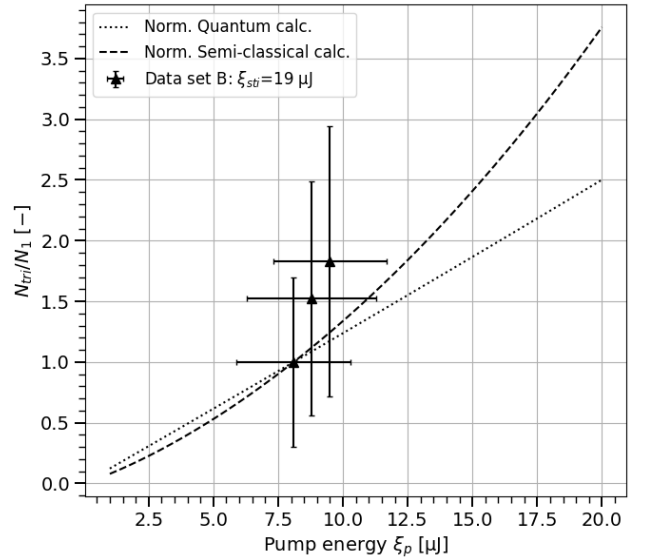

Fig. 4. Normalized flux of the generated photons as a function of the pump energy.

In a second step, the absolute magnitude of the generated flux is considered, where $T_F$ is taken at 11% that corresponds to the average value of the observed range. The corresponding curve is shown Fig. 5 where the number of generated triplets is plotted as a function of the input modes energy product $\xi_p\xi_{sti}$. Fig. 5 shows that we were able to generate from 2.2 triplets/s at $\{\xi_p = 19.3\ \mu J, \xi_{sti} = 3.04\ \mu J\}$ to 11.6 triplets/s at $\{\xi_p = 19.3\ \mu J, \xi_{sti} = 11.2\ \mu J\}$, the later one corresponding to 1.16 triplets/pulse since the repetition rate equals 10 Hz. This nearly verifies the criterion $N_{triplets} < 1$ triplets/pulse and undeniably attests the accuracy of true coincidence measurements for the other values. These results are consequently the first experimental demonstration of the non-seeded photons time-correlation in a third-order difference frequency mixing to the best of our knowledge. Figure 5 also indicates that the semi-classical model aligns perfectly with the measurements whereas the quantum model deviates up to nearly one order of magnitude. Because normalizing adjusts the model to the first experimental data point, the deviation here is higher and accurate compared to Fig. 3 and Fig. 4. At this point, there is no doubt regarding the performance of the semi-classical model compared to the non-linear momentum quantum model.

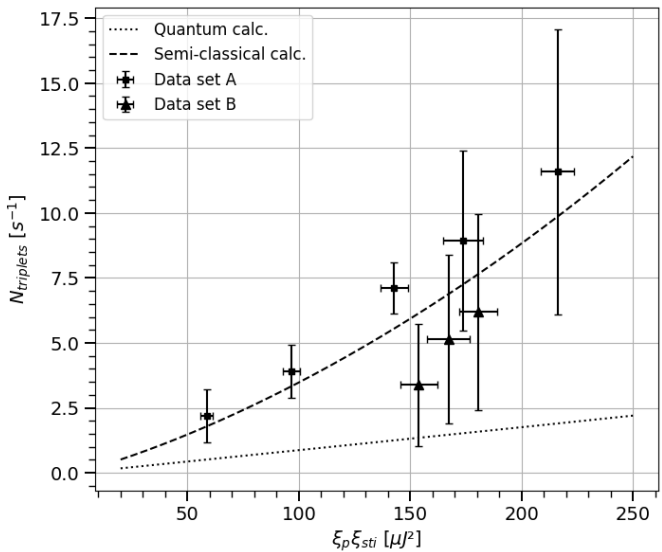

Fig. 5. Absolute magnitude of the flux of the generated photons as a function of the product of the stimulation and pump energies. The transfer function is taken at $T_F = 11\%$.

As a conclusion, using a configuration of difference-frequency generation governed by the third-order electric susceptibility, which corresponds to a photon-triplet generation stimulated over one mode of the triplet, we were able to generate and detect up to $N_{triplets}$ = 11.6 triplets/s from $N_{pump}$ = $5.2 \times 10^{14}$ pump-photons/s and $N_{sti}$ = $8.4 \times 10^{14}$ stimulation-photons/s. The corresponding quantum efficiency relative to the pump photons is then $N_{triplets} / N_{pump}$ = $2.23 \times 10^{-14}$ and the quantum parametric gain amplification $N_{triplets} / N_{sti}$ = $1.38 \times 10^{-14}$. These measurements were performed using a coincidence protocol, which was a "tour de force" according to the very small level of events and the very low repetition rate. The other strong point is the significant agreement of the semiclassical approach we have proposed compared to the quantum modelling. Next step will be the achievement of squeezing and entanglement measurements of same type than those performed for second-order SPDC [11,12]. Actually, in both cases photon-pairs are considered, knowing that here the photon-pair is part of the photon-triplet, *i.e.* two of the three photons. Such a goal will require to use a high repetition rate for the pump and stimulation lasers. Longer term and ambitious objective will be the spontaneous photon-triplet generation, which corresponds to a third-order spontaneous-parametric down-conversion (SPDC).

**Acknowledgments.** The authors acknowledge support from Program QuanTEdu-France n° ANR-22-CMAS-0001 France 2030 of the PhD fellowship of Gaspar Mougin-Trichon.

**Disclosures.** The authors declare no conflicts of interest.

**Data availability.** Data underlying the results presented in this paper are not publicly available at this time but may be obtained from the authors upon reasonable request.

**Supplemental document**. No supplement.